\documentclass[usenatbib]{basi}
\usepackage[british]{babel}
\usepackage[varg]{txfonts}
\usepackage{rotating}
\usepackage{dcolumn}
\begin{document}
\title[USS radio sources]{Optical observations of Ultra Steep Spectrum radio sources}
\author[George and Stevens]%
       {Samuel~J.~George$^{1,2}$\thanks{email: \texttt{sgeorge@mrao.cam.ac.uk}} and
       Ian~R.~Stevens$^{2}$\\
       $^1$Astrophysics Group, The Cavendish Laboratory, JJ Thomson Avenue, University of Cambridge, \\ 
           Cambridge, CB3 0HE, UK\\
       $^2$Astrophysics and Space Research Group, School of Physics and Astronomy, University of Birmingham, \\
           Birmingham, B15 2TT, UK 
       }

\pubyear{2011}
\volume{39}
\pagerange{\pageref{firstpage}--\pageref{lastpage}}

\date{Received 2011 July 22; accepted 2011 September 23}

\maketitle
\label{firstpage}

\begin{abstract}
In this paper we present follow-up optical observations of Ultra Steep Spectrum 
sources that were found by matching 150~MHz GMRT sources with either the 74~MHz 
VLSS or the 1400~MHz NVSS. These sources are possibly high-redshift radio galaxies 
but optical identification is required for clarification. 
The follow-up observations were conducted with the Liverpool Telescope; 
in all cases no sources are detected down to an R magnitude of $\sim$23. 
By applying models and using the $K-z$ relation we are able to suggest that these sources 
are possibly at high redshift. We discuss how 2m class telescopes can help with the 
identification of HzRGs from large-scale, low-frequency surveys. 
\end{abstract}


\section{Introduction}\label{s:intro}

Radio surveys play a crucial role in the study of active galactic 
nuclei (AGN) and they have several advantages over other methods. 
The emission tends to be much more powerful (than at other wavelengths). 
The most powerful sources at highest redshifts pin-point the most massive 
and luminous galaxies at such redshifts and crucially they are not affected 
by absorption by dust. In the local Universe, high-power 
radio galaxies are found in lower-density environments than low-luminosity 
radio galaxies. This means that powerful radio galaxies could serve as efficient 
probes of moderate redshift galaxy groups and poor clusters. 

Radio sources can be detected out to high redshifts ($z>5$). Empirically,
it appears that the majority of high-redshift radio galaxies (HzRGs) tend to
exhibit steeper radio spectra compared to lower redshift objects 
\cite[][]{debreuck2000, Miley2008}. 
The origin of this is probably due to radiative ageing of the electron energy 
distribution, essentially the higher energy electrons radiate faster. 
Part of this relation may be due to a Malmquist bias; if a sample of objects
is flux density limited then the observer will see an increase in the average
luminosity with distance (this is due to the less luminous
sources at large distances not being detected). 
It is likely that the higher gas densities in the galaxy 
environments at high redshifts play a role \cite[][]{Miley2008}.
The larger densities constrict the development of radio
bubbles from the AGN. 

We know that the faint radio population is a mixture of several types of 
objects including faint AGN, normal spirals and ellipticals, and 
starburst galaxies.  In general, very little is understood about the 
relative importance of the different classes of sources. Reasons for 
this include that the faint radio source samples are small, and the 
optical follow-up is incomplete (only 20$\%$ of sources have 
spectroscopic follow-up). With the spectroscopic redshifts a radio 
luminosity function (RLF) can be derived. The space density of 
sources increases with $z$ up to $z =2$ \cite[][]{Wall2005}.
\cite{Jarvis2001} found a constant comoving space density 
between $z \approx 2.5$ and $z \approx 4.5$. The main 
problem for establishing the high-$z$ RLF is the small number of $z > 2$ 
radio galaxies. To determine this more accurately one needs to determine 
a method for detection of these radio galaxies. A very successful method 
for finding HzRGs is to construct a filtered 
survey and this can easily be achieved by picking sources which are 
Ultra Steep Spectrum (USS) sources \cite[][]{Miley2008}. 

Having identified candidate HzRGs via their radio spectrum it is then
necessary to identify the galaxies in the optical/infrared wavelengths, and 
subsequently determine the redshifts. Bright, nearby sources, are discarded by 
comparing positions with wide-field shallow surveys at optical and 
infrared wavebands. Next, follow-up observations in optical and infrared wavelengths 
are undertaken. In general the near-IR $K$-band is the most effective 
and up to $94\%$ of sources have been detected down to $K = 22$ for sources 
with $S_{1.4GHz} < 50$~mJy  \cite[]{debreuck2002}. 
Once successful detections are made follow-up spectroscopic observations 
are made, allowing redshifts to be determined. HzRGs seem to have a relatively 
small scatter in a plot of $K$-band magnitude versus $z$,
and they seem to be among the most luminous galaxies at high redshift \cite[]{debreuck2002}. 
Apart from HzRGs there are a number of other sources
that exhibit steep spectral indices. These include, but are not limited
to, fossil radio galaxies, cluster halos and pulsars \cite[]{Parma2007,Manchester2005}.

In this paper we present follow-up observations of 3 USS sources 
determined by Giant Metrewave Radio Telescope (GMRT) observations 
at 150~MHz. The follow-up optical observations are conducted with the 
2m Liverpool Telescope using RATCam with 3 bands ($RIG$).

\section{USS Sources from the GMRT}\label{s:gmrt}

\begin{table}
\begin{center}
\caption[USS sources]{USS sources taken from  \cite{george2008}. All sources have a spectral index $\alpha>1.25$ in either the range 74--150~MHz or 150--1400~MHz.} \smallskip
\begin{tabular}{lccl}
\hline\\[-3mm] 
Name & $\alpha_{74}^{150}$ & $\alpha_{150}^{1400}$ & Surveys / Detections\\
     &&&\\ \hline 
GMRT J033152.9$-$100843 & $2.5$ & $0.8$ &  VLSS $\surd$, GMRT $\surd$, NVSS $\surd$ \\
GMRT J033216.0$-$083940  & -- & $>1.26$ & VLSS $\times$, GMRT $\surd$, NVSS $\times$\\
VLSS J033315.9$-$085119  & $>2.5$&-- & VLSS $\surd$, GMRT $\times$, NVSS $\times$ \\
GMRT J033318.7$-$085227 & -- & $>1.26$ & VLSS $\times$, GMRT $\surd$, NVSS $\times$\\
GMRT J033326.6$-$084205 & -- & $>1.42$ & VLSS $\times$, GMRT $\surd$, NVSS $\times$\\
\hline\\[-3mm] 
\end{tabular}
\label{uss}
\end{center}
\end{table}

\cite{george2008} presented a catalogue of 
113 sources at 150~MHz with the GMRT with a flux density limit of 
18.6~mJy (6$\sigma$). At $150$~MHz the GMRT has a spatial 
resolution approaching $20''$, a positional accuracy of a point 
source better than $4''$, and for an observation of 8 
hours a theoretical sensitivity of $\sim 1$~mJy beam$^{-1}$. 
The total area covered by the GMRT 
data in \cite{george2008} is $\sim 3$~deg$^{2}$.
Table \ref{uss} lists the 5 candidate USS sources 
identified by this survey both by their presence and absence at $150$~MHz. 
These sources were found by comparing the GMRT sources with both the 
National Radio Astronomy Observatory Very Large Array Sky 
Survey (NVSS) and the VLA Low-frequency Sky Survey (VLSS).
The NVSS was made at 1.4~GHz with a spatial resolution of $45''$ 
and a limiting source brightness of about 2.5 mJy beam$^{-1}$ \cite[]{condon1998}.
The VLSS was observed at 74~MHz with a spatial resolution of $80''$ and a typical 
source detection limit of $700$~mJy \cite[]{Helmboldt2008}. 
After the GMRT sources are matched with the VLSS and the NVSS the 
spectral indices $\alpha$ ($S \propto \nu^{-\alpha}$) are determined. 
USS sources are determined as having either $\alpha_{74}^{150}>1.25$ 
or $\alpha_{150}^{1400}>1.25$. This is similar to the condition 
applied by \cite{debreuck2000}; however they determined the 
spectral index between $325$ and $1400$~MHz
As discussed in the introduction, a steep radio spectral 
index does not automatically imply a high-redshift object 
and follow-up observations are necessary to confirm the 
nature of these sources.

For these objects we have investigated their nature using a 
variety of virtual observatory tools. None have been previously 
detected at optical wavelengths with the Digitized Sky Survey (DSS) or infrared 
wavelengths with either the Two Micron All Sky Survey (2MASS) or the
Infrared Astronomical Satellite (IRAS). 
From the DSS maps no source can be seen down to the completeness level 
($R$-band $\approx$ 20). From the 2MASS maps no source 
can be seen within twice the position errors of our GMRT observations. 
Based on this we determine that the source must be fainter 
than $J \sim 16$ and $H \sim 15.5$. Unfortunately, none of these 
sources are covered by UKIRT Infrared Deep Sky Survey (UKIDDS) 
or the Sloan Digital Sky Survey (SDSS). 

Three of the sources from \cite{george2008} were chosen for follow-up 
observations. These sources are listed in Table \ref{USS_LT} and the  
results of the virtual observatory search are given below:

\begin{enumerate}
\item GMRT J033152.9$-$100843  - the nearest source in the $J$-band 
(from 2MASS) is $41''$ away with a magnitude of 16.8.

\item VLSS J033315.9$-$085119 - the nearest source in the $J$-band 
(from 2MASS) is $47''$ away with a magnitude of 15.7.
 
\item GMRT J033326.6$-$084205 - the nearest source in the $J$-band 
(from 2MASS) is $16''$ away with a magnitude of 12.23; 
this may end up being a confusing source.
\end{enumerate}

\section{Optical observations}\label{s:observations}

The Liverpool Telescope is a $2$m unmanned fully robotic 
telescope at the Observatorio del Roque de Los Muchachos on 
the Canary island of La Palma \citep{Steele2004}. 
For our observations we chose to use the RATCam instrument which 
consists of a $2048\times2048$ pixel EEV CCD42-40, back-illuminated 
broadband chip with a pixel size of $13.5\mu$m. These observations 
used filters: SDSS-$I$ (693-867~nm), SDDS-$R$ (556-689~nm) and 
SDSS-$G$ (410-550~nm). Standard fields are based on the \cite{Landolt19992} 
series of standards. They are spaced every few hours of RA and are observed 
in all bands. Basic instrumental reductions are applied to all RATCam 
images before the data are passed to users. This includes bias subtraction, 
trimming of the overscan regions and flat fielding. The data were obtained 
during photometric conditions with exposures of $400$s ($I$), 
$150$s ($R$), $50$s ($G$). Full observational details can be 
found in Table \ref{LT_results}. Zero-point calibration was 
provided by observations of Landolt stellar and star fields \cite[]{Landolt19992}.

The observations took place between 17th and 29th August 2008 
in photometric conditions. These observations are essentially 
a pilot for detecting HzRGs with the GMRT and 2m class robotic telescopes.

\section{Results and discussion}\label{s:results}

\begin{figure}

\centerline{\includegraphics[scale=0.62]{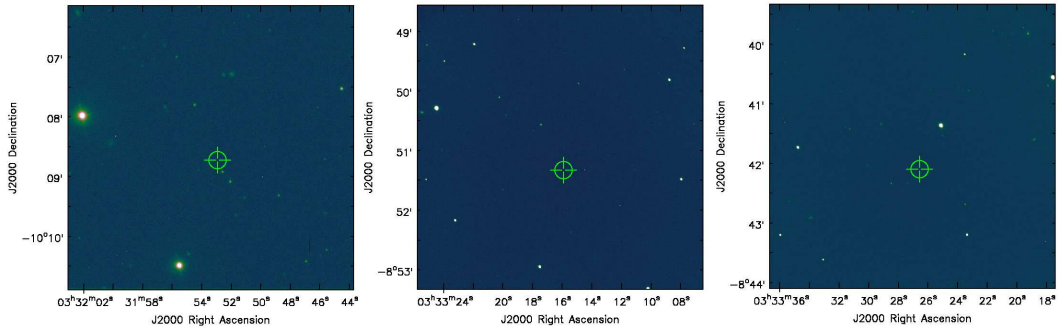} \qquad
}

\caption[GMRT LT $I$ images]{$I$ band images of GMRT J033152.9$-$100843 (left), 
VLSS J033315.9$-$085119 (middle) and GMRT J033326.6$-$0842 (right) taken with the Liverpool Telescope). In all cases no sources are found at the position of the USS source. The field of view of RATCAM is $4.6'$. \label{LT_1}
}
\end{figure}

Our Liverpool Telescope observations go much deeper than 
the DSS, theoretically reaching (for a 3$\sigma$ point source)  
$G = 22.8$, $I = 23.3$, $R = 23.5$ with these observations 
approaching these estimates. The sensitivity of our observations 
are listed in Table \ref{LT_results}.
Fig. \ref{LT_1} shows $R$-band 
images of GMRT J033152.9$-$100843, VLSS J033315.9$-$085119 and 
GMRT J033326.6$-$084205. In all the observing bands no source 
can be seen at the location of the radio source. 
Deeper high-resolution multi-band 
optical observations are required to detect the optical counterparts of 
these sources and especially to resolve and characterise them.

\begin{table} 
\begin{center}
\caption[Liverpool Telescope USS follow-up targets]{The USS radio sources followed up with the
Liverpool Telescope. The source name, the object position, the radio flux densities at 74~
MHz, 150~MHz and 1400~MHz and the source angular size (for GMRT sources at 150~MHz for VLSS at 74~MHz)
are listed. \label{USS_LT}}\smallskip 
\begin{tabular}{cccccccc}
\hline\\[-3mm] 
 Source Name& RA & DEC &$S_{1400}$ &$S_{150}$ & $S_{74}$ & Size\\
           & (J2000.0)&(J2000.0)&(mJy) &(mJy)         & (mJy)    & ($''$)\\ 
\hline\\[-3mm] 
GMRT J033152.9$-$100843& 03 31 52.97& $-$10 08 43.72&37.0  &142.2  &950 & 95.8 \\  
VLSS J033315.9$-$085119& 03 33 15.89& $-$08 51 19.90&$<2.5$&$<18.0$ &970 & $\approx$130\\ 
GMRT J033326.6$-$084205& 03 33 26.61& $-$08 42 05.86&$<2.5$& 56.4  &$<100$ & 39.1\\ 
\hline\\[-3mm] 
\end{tabular}
\end{center}
\end{table}

\begin{table}
\caption[Liverpool Telescope RATCam imaging results]{Results from Liverpool Telescope RATCam imaging. Given is the source name, observation date, the filter used, the exposure time and the observed limiting magnitude. 
\label{LT_results}}
\begin{center}
\begin{tabular}{llllll}
\hline
Source Name &Obs. Date &Filter & Exposure & Mag. \\ 
            & &      & (secs)     &Limit\\ 
\hline
GMRT J033152.9$-$100843   & 2008-08-29 & $I$ &400 &23.0\\ 
                          & 2008-08-29 & $R$ &150 &23.2\\ 
                          & 2008-08-29 & $G$ &50  &22.6\\ 
VLSS J033315.9$-$085119   & 2008-08-17 & $I$ &400 &22.5\\ 
                          & 2008-08-17 & $R$ &150 &22.5\\ 
                          & 2008-08-17 & $G$ &50  &22.0\\
GMRT J033326.6$-$084205   & 2008-08-17& $I$ &400 &22.1\\ 
                          & 2008-08-17& $R$ &150 &23.0\\ 
                          & 2008-08-17 & $G$ &50 &22.0\\ 
\hline
\end{tabular}
\end{center}
\end{table}

We have presented $GRI$-band magnitude limits on sources
with steep radio spectra. For these steep-spectrum sources the most
commonly used (and most useful) band is the $K$-band. We will now 
relate our limits to these bands. \cite{Miley2008} have presented an overview 
of the optical/infrared properties of HzRGS, showing the long-established 
correlation between the $K$-band magnitude and redshift -- the $K-z$ relationship.

\cite{Bryant2009} discuss the $K-z$ correlation for a large sample
of USS-selected radio galaxies using data from 
the $408$-MHz Revised Molonglo Reference Catalogue and the $843$-MHz
Sydney University Molonglo Sky Survey (the MRCR-SUMSS sample), and the 
NVSS at $1400$\,MHz. Fitting the
data, these authors find a correlation of the form

\begin{equation} 
K = 17.75 + 3.64 \log_{10}(z) 
\end{equation} 

\noindent where the $K$-band luminosity is measured with a $4''$ aperture,
and the relation applies for objects with $1 < z < 4$, and with a dispersion
of around 0.7\,mag on the relationship.  Using spectral evolution
models of \cite{Rocca-Volmerange2004}, \cite{Bryant2009} find
the properties of the galaxies consistent with being elliptical
galaxies with masses in the range $10^{11}-10^{12}$\,M$_\odot$.

\cite{debreuck2002} have discussed optical/near-IR imaging of 128
USS sources and noted that while 94$\%$ of the radio sources were
identified down to a magnitude of $K=22$, only around 50$\%$ were
identified down a magnitude of $R=24$. \cite{debreuck2002} found
that there was a large range in the $R-K$ colour of the USS sources,
with some sources having values of $R-K$ in the range of 3--6, while
for others, there are only limits with implied values of $R-K>6$.

Using data from FIRST (Faint Images of the Radio Sky at 20\,cm), 
\cite{Bouchefry2009a,Bouchefry2009b} has presented $R-K$ vs $K$ and $I-K$ vs
$K$ colour-magnitude relationships for the sample of radio
galaxies. The author notes that while there is a similar correlation 
between the $R-K$ and $I-K$ colours, there is also a larger scatter seen. 
Consequently, $R$ and $I$ band imaging is capable of detecting
counterparts to USS sources, though the variation in the $R-K$ and
$I-K$ colours means that the success rate is going to be variable. In
this paper we have presented $R$-band limits of $\sim 23$\, mag, and
these will correspond to $K$-band limits in the general range of
$17-20$. By using Fig. 12 of \cite{Miley2008}
and assuming that the galaxy mass is $10^{11} M_{\odot}$ 
we are able to suggest that the sources are at $z > 0.5$. 

\section{Summary}\label{s:conclusions}

USS sources were identified with the GMRT and we have 
conducted a virtual observatory search and follow-up 
observations. The observations of 3 USS sources 
with the Liverpool Telescope are ultimately inconclusive, 
with no detection in any of the 3 wavebands. Observations 
at infrared (in particular $H$ and $J$ bands) are required. 
We have presented $R$-band limits of $\sim 23$\, mag, and
these correspond to $K$-band limits in the general range of
$17-20$. By using correlations between $K$-band magnitude and 
redshift we can tentatively suggest that these are sources with 
$z>0.5$. 

The GMRT is well suited to complete a large sky survey of 
the low-frequency radio source population, going far deeper 
than has currently been explored with past surveys. If this 
is combined with archival observations and new targeted 
observations then this will enable the detection of previously 
undetected high-$z$ objects amongst other sources. The ongoing
TIFR GMRT Sky Survey (TGSS) - the radio continuum survey 
at $150$\,MHz using GMRT covering $\sim 30\,000$deg$^{2}$ 
offers exciting prospects for the detection 
of many sources with interesting spectral indices. In the case of 
HzRG candidates once they have been identified via their spectral index 
it is crucial that deep optical/infrared follow-up is done. 
When there is no previous optical coverage we have 
shown that short $2$m-class observations 
allow for preliminary exploration of sources. In the case 
of large-scale surveys this will be of great importance to
filter the un-interesting sources.

\section*{Acknowledgements}

We thank the GMRT staff who have made these observations possible. The
GMRT is run by the National Centre for Radio Astrophysics
of the Tata Institute of Fundamental Research. In particular,
we would like to thank Ishwara Chandra C. H. for his input
during the observations and data reduction. The Liverpool Telescope is operated 
on the island of La Palma by Liverpool John Moores University in the Spanish 
Observatorio del Roque de los Muchachos of the Instituto de Astrofisica de Canarias 
with financial support from the UK Science and Technology Facilities Council. 
This paper makes use of the CubeHelix colour scheme of \cite{green2011}.
We thank the referee for a detailed and helpful report
on the paper.


\label{lastpage}
\end{document}